\documentstyle{article}
\newcommand{\newsection}{    
\setcounter{equation}{0}
\section}
\def\appendix#1{
\addtocounter{section}{1}
\setcounter{equation}{0}
\renewcommand{\thesection}{\Alph{section}}
\section*{Appendix \thesection\protect\indent #1}
\addcontentsline{toc}{section}{Appendix \thesection\ \ \ #1}
}

\newcommand{\ov}[1]{\overline{#1}}

\def\be{\begin{equation}}
\def\ee{\end{equation}}
\newcommand{\beq}{\begin{equation}}
\newcommand{\eeq}{\end{equation}}
\newcommand{\bea}{\begin{eqnarray}}
\newcommand{\eea}{\end{eqnarray}}

\newcommand{\th}{\theta}

\newcommand{\vph}{\varphi}

\newcommand{\eps}{\varepsilon}

\def\e{{\,\rm e}\,}

\renewcommand{\d}{{{\partial}}}

\begin{document}
\topmargin 0pt
\oddsidemargin 5mm
\headheight 0pt
\headsep 0pt
\topskip 9mm

{{}\hfill SMI-Th--98/23}

\begin{center}
\vspace{26pt}
{\large \bf {AdS/CFT correspondence on torus}}
\vspace{26pt}

{\sl L.\ Chekhov} 
\footnote{email: chekhov@mi.ras.ru}
\\ 
\vspace{6pt} 
Steklov Mathematical Institute\\ Gubkin 
st.\ 8, 117966, GSP-1, Moscow, Russia\\

\end{center}
\vspace{20pt} 
\begin{center}
{\bf Abstract}
\end{center}

\noindent
The AdS/CFT correspondence is established for
the case of AdS$_3$ space 
compactified on a filled rectangular torus with the CFT field on
the boundary.


\newsection{Introduction}
The AdS/CFT correspondence~\cite{Maldacena,Polyakov,Witten} 
is currently under study in various respects.
One may check its validity for the case of interacting fields~\cite{inter,HF}
(multi-particle scattering in gravitation theory, etc.)
but it would be also interesting to check whether it holds in
cases where the space--time geometry is more involved than 
the spherical one. 

In~\cite{Bonelli}, it was
proposed, starting from a two-dimensional (compact) manifold $M$,
to consider a theory on the space $M\times {\bf R}_+$ endowed with the
AdS metric. From the topological standpoint, this, however, results
in a singularity as $s\in{\bf R}\to+\infty$ (because $M$ is not
necessarily simply connected), and one must impose an additional
condition on the fields of the theory (the fast decrasing at infinity)
in order to make the field configuration smooth.

In this paper, we consider the case of AdS$_3\times S^{d+1}$ space 
and confine ourselves to the massless case (the case 
where the internal degrees of freedom w.r.t.\ the internal compact group
$S^{d+1}$, \ $d=2,3,\dots$, are swtiched off). However, this is only
for the integrity of the paper, and all the main stages
of our consideration can be generalized to the massive modes of 
AdS$_n\times S^{d+1}$ space. We also consider the case of a homogeneous
compactified AdS$_3$ manifold without (topological) singularity in the
interior (we just have no boundary as $s\to\infty$; note also that
our calculations technically resemble the bulk calculations of~\cite{HF}).
We show that the classical scalar field theory on the AdS$_3$ manifold
gives us the appropriate quantum correlators.

\newsection{Geometry of AdS$_3$ manifolds}

The group $SL(2,{\bf C})$ of conformal transformations of the complex
plane admits the continuation to the upper half-space ${\bf H}_3^+$
endowed with the constant negative curvature (AdS$_3$ space).
In the Schottky uniformization picture, Riemann surfaces of higher genera
can be obtained from ${\bf C}$ by factorizing it over a finitely
generated free-acting discrete subgroup $\Gamma\subset SL(2,{\bf C})$.
Therefore, we can continue the action of this subgroup to the 
whole AdS$_3$ and, after factorization, obtain a three-dimensional manifold
of constant negative curvature (an AdS$_3$ manifold)
whose boundary is (topologically)
a two-dimensional Riemann surface~\cite{Manin,Sel}.

We consider the simplest case of a genus one AdS$_3$-manifold, which
can be obtained if we identify
\be
(w,{\ov w},s)\sim (qw,{\ov q}{\ov w},s|q|),
\label{a1}
\ee
where $q=\e^{a+ib}$ is the modular parameter, $a,b\in {\bf R},\ a>0$,
$w,\ov w=x+iy,\,x-iy$ are 
the coordinates on $\bf C$, and $s$ is the third coordinate on AdS$_3$.

Adopting the AdS/CFT correspondence approach, we should first regularize
expressions in order to make them finite. In the AdS$_n$ case with the
(brane or black hole) singularity 
at infinity, this was done by setting the
boundary data on the $\eps$-plane~\cite{Polyakov} 
rather than at infinity (zero plane).
However, in our case, we cannot take an $\eps$-plane because it is not
invariant w.r.t.\ transformations (\ref{a1}). Instead, we can set
the boundary data on the $\eps$-{\it cone}---the set of points 
$$
z/r=\eps, \quad r^2=w\ov w+s^2. 
$$
Given the boundary data on this cone, we
fix the problem setting---the Laplace equation then has a unique solution
(the Dirichlet problem on a compact manifold)
and we do not need to introduce additional constraints (as in the case
of AdS$_3$). 

Geometrically, performing the $\eps$-cone 
regularization and factorization over
the group $\Gamma$ of transformations (\ref{a1}), we obtain the filled
torus whose boundary is associated with the two-dimensional manifold
(the torus) on which the CFT dwells. The ``center'' of the torus is now
the only closed geodesic of the length $\log|q|$
(the image of the vertical half-line $z=\ov z=0$), 
and the AdS-invariant
distance $r$ from this geodesic to the image of the $\eps$-cone is constant,
$\cosh r=1/\eps$. 

\newsection{Scalar field on AdS$_3$ in spherical coordinates}

In order to operate with the cone geometry, it is convenient to
reformulate the free-field problem on
AdS$_3$ in the spherical coordinates $r\equiv\e^\tau,\th,\vph$
in which relations (\ref{a1}) read
\be
(\tau,\th,\vph)\sim (\tau+ma, \th, \vph+mb+2\pi n),\qquad m,n\in \bf Z.
\label{a2}
\ee
On the cone $\sin\th=\eps$, the usual toric periodic conditions
are imposed on fields depending on the two-dimensional variables $\tau,\vph$.

Spherical coordinates are standard,
\be
\begin{array}{ll}
s=r\cos\th &\d_s=\cos\th\d_r-\frac{1}{r}\sin\th\d_\th\\
y=r\sin\th\sin\vph &\d_y=\sin\th\sin\vph\d_r+\frac{1}{r}\cos\th\sin\vph\d_\th
+\frac{1}{r\sin\th}\cos\vph\d_\vph\\
x=r\sin\th\cos\vph &\d_x=\sin\th\cos\vph\d_r+\frac{1}{r}\cos\th\cos\vph\d_\th
-\frac{1}{r\sin\th}\sin\vph\d_\vph  
\end{array}
\ee
The action of the massless scalar field~$\Phi$ on AdS$_3$ is
\bea
&{}&\int\frac{dx\,dy\,ds}{s^3}\bigl\{s^2\d_s\Phi\d_s\Phi+s^2(\d_x\Phi\d_x\Phi
+\d_y\Phi\d_y\Phi)\bigr\}=\nonumber\\
&{}&\qquad=\int\,r\tan\th dr\,d\th\,d\vph\left\{\d_r\Phi\d_r\Phi
+\frac{1}{r^2}\d_\th\Phi\d_\th\Phi
+\frac{1}{r^2\sin^2\th}\d_\vph\Phi\d_\vph\Phi\right\}=\nonumber\\
&{}&\qquad=
\int\,\tan\th d\tau\,d\th\,d\vph\left\{\d_\tau\Phi\d_\tau\Phi
+\d_\th\Phi\d_\th\Phi
+\frac{1}{\sin^2\th}\d_\vph\Phi\d_\vph\Phi\right\}
\label{action}
\eea
It admits the variable separation:
\be
\Phi(\tau,\th,\vph)=
\sum_{k,m\in {\bf Z}}^{}\Phi_{l}(\tau)Y_{k,m}(\sin\th)X_m(\vph),
\ee
where
\be
X_m(\vph)=\e^{im\vph}; \qquad \d_\vph^2X_m=-m^2X_m,
\label{vph}
\ee
and for the function $Y_{k,m}(\sin\th)$, the eigenvalue problem arises
\be
\frac{\cos\th}{\sin\th}\d_\th\left(\frac{\sin\th}{\cos\th}
\d_\th Y_{k,m}(\sin\th)\right) -\frac{m^2}{\sin^2\th}Y_{k,m}(\sin\th)
=\lambda_k Y_{k,m}(\sin\th),\quad \lambda_k\ge0.
\ee
We substitute $\rho$ for $\sin\th$ and consider the problem on the
interval $1-\eps\ge\rho\ge0$ with the regularity condition at $\rho=0$,
\be
Y''_{k,m}(\rho)+\frac{1}{\rho}Y'_{k,m}(\rho)
-\frac{m^2}{\rho^2(1-\rho^2)}Y_{k,m}(\rho)=\frac{\lambda_k}{1-\rho^2}
Y_{k,m}(\rho).
\label{e1}
\ee
The equation for the radial coordinate,
\be
\Phi''_{k}(\tau)+\lambda_k\Phi_{k}(\tau)=0,
\label{e2}
\ee
determines the values of $\lambda_k$ consistent with
the periodicity conditions
\be
\Phi(\tau+\log|q|,\th,\vph+\arg q)=\Phi(\tau,\th,\vph),
\label{e3}
\ee
where $q$ is the modular parameter of the torus,
\be
q=\e^{a+ib},\quad a,b\in{\bf R}.
\label{e4}
\ee

Now Eq.~(\ref{e2}) can be easily solved, $\lambda_k=-mb/a+2\pi k/a$,
and Eq. (\ref{e1}) can
be reduced to the standard hypergeometric equation whose
general solution that is regular at $\th=0$ yields
\bea
\Phi(\tau,\th,\vph)&=&\sum_{m,k\in{\bf Z}}^{}\e^{im\vph}\,
\e^{i\bigl[-\frac{mb}{a}+\frac{2\pi k}{a}\bigr]\tau}
\times C_{k,m}\times Y_{k,m}(\sin\th),
\nonumber\\
Y_{k,m}(\sin\th)&=&[\sin\th]^{|m|}\,
{}_2F_1\left(\frac{|m|}{2}+i\frac{-mb+2\pi k}{2a},
\frac{|m|}{2}-i\frac{-mb+2\pi k}{2a}; |m|+1; \sin^2\th\right),
\label{e5}
\eea
where ${}_2F_1(a,b;c;z)$ is the hypergeometric series,
$$
{}_2F_1(a,b;c;z)\equiv\sum_{k=0}^{\infty}\frac{(a)_k(b)_k}{(c)_k\cdot k!}z^k,
\qquad (a)_k=\prod_{i=0}^{k-1}(a+i),
$$
and $C_{k,m}$ are the mode amplitudes.

Expression (\ref{e5}) is singular at $z=1$ and we must find its asymptotic
behavior for $z=1-\eps$. The standard formula
\bea
{}_2F_1(a,b;c;z)&=&\Gamma\left[\begin{array}{ll} c,& c-a-b\\
                                                c-a,& c-b \end{array}\right]
{}_2F_1(a,b; 1+a+b-c;1-z)+
\nonumber\\
&{}&+\Gamma\left[\begin{array}{ll} c,& a+b-c\\
                                   a,& b \end{array}\right]
(1-z)^{c-a-b}{}_2F_1(c-a,c-b; 1+c-a-b;1-z)
\label{e6}
\eea
works 
for $c-a-b\not\in{\bf Z}$, while for $c-a-b\in{\bf Z}$ another
exact relation holds (see, e.g., formula 7.3.1.31 from~\cite{BPM}),
\bea
\Bigl.{}_2F_1(a,b;c;z)\Bigr|_{c=a+b+m}&=&
\Gamma\left[\begin{array}{ll} m,& a+b+m\\
                              a+m,& b+m \end{array}\right]
\sum_{k=0}^{m-1}\frac{(a)_k(b)_k}{k!(1-m)_k}(1-z)^k-
\nonumber\\
&{}&-\Gamma\left[\begin{array}{c} a+b+m\\
                              a,\ b \end{array}\right](z-1)^m
\sum_{k=0}^{\infty}\frac{(a+m)_k(b+m)_k}{k!(m+k)!}(1-z)^k\times
\nonumber\\
&{}&\qquad\times\bigl[\log(1-z)-\Psi(k+1)-\Psi(k+m+1)+\nonumber\\
&{}&\qquad\qquad
+\Psi(a+k+m)+\Psi(b+k+m)\bigr].
\label{e7}
\eea
Here $\Psi(x)$ is the logarithmic derivative of the $\Gamma$-function.
In the massless case, we need only the case $m=1$ in (\ref{e7})
and are interested in the asymptotic behavior as
$1-z\simeq\eps\to+0$.

Coefficients $C_{k,m}$ determine the boundary values of the field $\Phi$.
Then, for action (\ref{action}), we obtain
\bea
&{}&\int_0^{a}d\tau\int_0^{2\pi}d\vph\int_{\sin\th=0}^{\sin\th=1-\eps}d\th\,
\frac{\sin\th}{\cos\th}\left\{\d_\tau\Phi\d_\tau\Phi+\d_\th\Phi\d_\th\Phi
+\frac{1}{\sin^2\th}\d_\vph\Phi\d_\vph\Phi\right\}=\nonumber\\
&{}&\qquad=\int_0^a\,d\tau\int_0^{2\pi}\,d\vph \Phi(\tau,1-\eps,\vph)
\sin\th\left.\frac{\d}{\d\sin\th}
\Phi(\tau,\sin\th,\vph)\right|_{\sin\th=1-\eps}.
\label{e8}
\eea
Keeping only the logarithmically divergent and finite parts as $\eps\to0$,
we obtain action (\ref{action})
in the mode expansion ($C^*_{k,m}\equiv C_{-k,-m}$):
\be
\sum_{k,m\in{\bf Z}}^{}|C_{k,m}|^2\left\{|m|-2\left(\frac{m^2}{4}+w^2\right)
\bigl[\log\eps+\Psi(|m|/2+1+iw)+\Psi(|m|/2+1-iw)-2\Psi(1)\bigr]\right\}.
\label{e9}
\ee
Here
\be
w=\frac{-mb}{2a}+\frac{\pi k}{a},\qquad q=\e^{a+ib}, \quad a,b\in \bf R.
\label{e10}
\ee

\newsection{Massive modes}

Including into consideration the ``internal'' (compact) degrees of 
freedom (assuming the compact manifold to be a sphere $S^{d+1}$), we obtain
an additional term in the initial action (\ref{action}),
\bea
&{}&\int\frac{dx\,dy\,ds}{s^3}\bigl\{s^2\d_s\Phi\d_s\Phi+s^2(\d_x\Phi\d_x\Phi
+\d_y\Phi\d_y\Phi)+l(l+d)\Phi^2\bigr\}=\nonumber\\
&{}&\qquad=\int\,\tan\th d\tau\,d\th\,d\vph\left\{\d_\tau\Phi\d_\tau\Phi
+\d_\th\Phi\d_\th\Phi
+\frac{1}{\sin^2\th}\d_\vph\Phi\d_\vph\Phi
+\frac{l(l+d)}{\cos^2\th}\Phi^2\right\}.\label{action1}
\eea
Separating the variables as above, we obtain that the 
conditions on the ``torus'' coordinates~$\vph$ and~$\tau$  are as 
in (\ref{vph}) and (\ref{e2}),
and only the equation for the
$\th$-component is changed,
\be
\frac{\cos\th}{\sin\th}\d_\th\left(\frac{\sin\th}{\cos\th}
\d_\th Y_{k,m}\right)-\frac{m^2Y_{k,m}}{\sin^2\th}-\lambda_kY_{k,m}-
\frac{l(l+d)Y_{k,m}}{\cos^2\th}=0.
\label{sph1}
\ee
Let us introduce a new quantity $\rho$,
\be
4\rho(\rho-1)=l(l+d), \qquad \rho>0.
\ee
The solution to (\ref{sph1}) that is regular at $\th=0$ is
\be
Y_{k,m}(\th)=(\cos\th)^{2-2\rho}(\sin\th)^{|m|}
{}_2F_1\left(\frac{|m|}{2}+1-\rho+iw, \frac{|m|}{2}+1-\rho-iw; |m|+1; 
\sin^2\th\right),
\label{Ykm1}
\ee
where $w=-\frac{mb}{2a}+\frac{\pi k}{a}$.

Important particular case is $d=2$ (AdS$_3\times S^3$)
where $\rho=l/2+1$ and we must use
formula (\ref{e7}) rather than (\ref{e6}) in order to find the asymptotic
behavior. In this case, the nonlocality is again encoded 
in the $\Psi$-function terms.

\newsection{Exact Green's function for massless modes}

Turn now to expression (\ref{e9}). The term in braces is the Fourier
transform of the Green's function of the boundary CFT field on torus. 
We are interested in the 
nonlocal contribution to Green's functions on torus coming from this
formula and disregard all local terms (which can be removed by 
a proper renormalizing procedure). First, note that the $m$-dependence in
(\ref{e9}) can be set analytic. Using the standard formulas
$$
\Psi(x)=\Psi(1-x)-\pi\cot\pi x\quad \hbox{and}\quad 
\Psi(1+x)=\frac{1}{x}+\Psi(x),
$$
we can rewrite the summand in (\ref{e9}) as follows:
\bea
&{}&
|m|-2\left(\frac{m^2}{4}+w^2\right)\bigl[\Psi\left(1+|m|/2+iw\right)
+\Psi\left(1+|m|/{2}-iw\right)-2\Psi(1)\bigr]=\nonumber\\
&{}&\quad=
-\left(\frac{m^2}{4}+w^2\right)
\bigl[\Psi\left(1+{m}/{2}+iw\right)
+\Psi\left(1+{m}/{2}-iw\right)+\bigr.\nonumber\\
&{}&\qquad
+\bigl.\Psi\left(1-{m}/{2}+iw\right)
+\Psi\left(1-{m}/{2}-iw\right)
-4\Psi(1)\bigr],
\label{d1}
\eea
where the $\cot$-terms cancel each other because of the symmetry $w\to-w$.

In what follows, it is useful to represent the $\Psi$-function 
using the formula
$$
\Psi(1+a)-\Psi(1+b)=\sum_{j=0}^{\infty}\left(\frac{1}{1+j+b}
-\frac{1}{1+j+a}\right).
$$
Now we aim at finding the Green's function of the Yang--Mills field 
insertions on the two-dimensional torus. First, note that the term
$\left(\frac{m^2}{4}+w^2\right)$ in front of the summand is nothing but
the Laplacian action on the Riemann surface, i.e., 
we obtain that the correlation (Green's)
function is
\bea
G(\tau,\vph)&=&
\frac{1}{4}\bigl(\d_\vph^2+\d_\tau^2\bigr)
\sum_{m,k\in \bf Z}^{}
\e^{im\vph+i\left[-\frac{mb}{a}+\frac{2\pi k}{a}\right]\tau}
\left[2\log\eps+\sum_{(\pm),\pm}
\Psi\bigl(1(\pm){m}/{2}\pm iw\bigr)-4\Psi(1)\right]=\nonumber\\
&=&
\frac{1}{4}\bigl(\d_\vph^2+\d_\tau^2\bigr)
\sum_{m,k\in \bf Z}^{}
\e^{im\vph+i\left[-\frac{mb}{a}+\frac{2\pi k}{a}\right]\tau}\times\nonumber\\
&{}&\qquad\times\left[2\log\eps+\sum_{(\pm),\pm}
\sum_{l=1}^{\infty}\left(\frac{4}{l}
-\frac{1}{l(\pm) \frac{m}{2}\pm i\bigl(-\frac{mb}{2a}
+\frac{\pi k}{a}\bigr)}\right)\right].
\label{d2}
\eea
The sum $\sum_{(\pm),\pm}$ in (\ref{d2}) means that we must take a sum over
four terms with all possible appearances of the ``$+$'' and ``$-$'' signs.
We distinguish between two appearances of $\pm$ signs by
taking one of them in parentheses.
The triple sum over $m$, $k$, and~$l$ is rather involved.
Note that constant terms are irrelevant to our discussion as they
produce only local contributions (like $\d_z\d_{\ov z}
\delta^2_{\Pi}(z,\ov z)$, where $\delta^2_{\Pi}(z,\ov z)$ is the 
double-periodic two-dimensional delta function). 

First, we take the sum over $k$,
\be
\sum_{k=-\infty}^{\infty}{\e^{2\pi ik\tau/a}\over l(\pm) \frac{m}{2}\mp
\bigl(\frac{imb}{2a}-\frac{i\pi k}{a}\bigr)}=f(\tau/a).
\label{d3}
\ee
The function $f(\tau/a)$ is periodic under the shift
$\tau\to\tau+1$ and
satisfy the functional equation
\be
\pm\frac{1}{2a}\d_{\tau/a}f(\tau/a)+\left(l(\pm)\frac{m}{2}
\mp\frac{imb}{2a}\right)f(\tau/a)=\delta^1_{\Pi}(\tau/a).
\label{d4}
\ee
The (unique) solution to (\ref{d4}) that is periodic in $\tau$ is a 
saw-tooth-like exponential curve 
$$
f(\tau/a)=A\e^{\chi\tau/a}\quad\hbox{for}\quad \tau/a\in (0,1),
$$
which is to be continued periodically to the whole $\bf R$. Equation (\ref{d4})
gives
$$
\chi=\mp 2a\bigl(l(\pm){m}/{2}\bigr)+imb \quad\hbox{and}\quad 
A=\frac{\pm 2a}{1-\e^{\mp2a\bigl(l(\pm)m/2\bigr)+imb}}.
$$
Therefore, the remaining nonlocal terms are combined into the sum
\be
\sum_{{m\in {\bf Z}\atop l=1}}^{\infty}\e^{im\vph-imb\tau/a}
\left[\hbox{const}-\sum_{\pm,(\pm)}^{}
\frac{\pm2a\e^{\mp 2\bigl(l(\pm){m}/{2}\bigr)\tau+imb\tau/a}}
{1-\e^{\mp2a\bigl(l(\pm)m/2\bigr)+imb}}\right],\quad 0\le\tau<a.
\label{d5}
\ee

Next sum to be taken is (see, e.g.,~\cite{AF})
\be
\sum_{m\in \bf Z}^{}\frac{\e^{mw}}{1-q^m\e^u}=
\frac{\theta(u-w|q)\theta'(0|q)}{\theta(w|q)\theta(u|q)},
\quad |q|>1,\ 0<\hbox{Re\,} w,\hbox{Re\,} u<\log|q|,
\label{d6}
\ee
where $\theta(w|q)$ is the standard antisymmetric theta
function $\theta_{11}(w|q)$, which reads in our convention
as follows:
$$
\theta_{11}(w|q)\equiv \sum_{n\in \bf Z}^{}q^{-\frac{(n+1/2)^2}2}
\e^{(w+\pi i)(n+1/2)}.
$$
We exhibit the dependence on the modular parameter as it can be varied as well.
In particular, we have both the $q$- and $\ov q$-dependences.
Substituting (\ref{d6}) in (\ref{d5}) and taking all sums
over $\pm$ signs, we obtain
\bea
&{}&2\hbox{Re\,}\sum_{l=1}^{\infty}\left[\hbox{[local]\,}
-2a\e^{2il\vph}{\theta\bigl(2ilb-\tau-i\vph|\e^{a+ib}\bigr)
\theta'\bigl(0|\e^{a+ib}\bigr)\over
\theta\bigl(\tau+i\vph|\e^{a+ib}\bigr)\theta\bigl(2ilb|\e^{a+ib}\bigr)}
\right.
\nonumber\\
\qquad\qquad\qquad\qquad &{}&\left.
+2a\e^{-2il\vph}{\theta\bigl(-2ilb-\tau-i\vph|\e^{a+ib}\bigr)
\theta'\bigl(0|\e^{a+ib}\bigr)\over
\theta\bigl(\tau+i\vph|\e^{a+ib}\bigr)\theta\bigl(-2ilb|\e^{a+ib}\bigr)}
\right].
\label{d7}
\eea

Let us introduce the complex variables 
$$
z=\tau+i\vph, \qquad \ov z =\tau-i\vph.
$$ 
Unfortunately, we are inaware of whether sums of type
(\ref{d7}) has been presented
in the literature.\footnote{As we hope, there must be such formulas in 
the general case enjoying nice properties of modular invariance.} \ 
However, we can proceed with an important particular case of a rectangular
torus. This corresponds to taking the limit $b\to0$, which 
must be nonsingular as follows from the problem setting.

In the asymptotic regime $b\to0$, the leading term, which may lead to 
a divergency in (\ref{d7}), vanishes because of the real part condition
resulting from the $w\to-w$ invariance,
$\lim_{b\to0}\theta(z+2ilb|q)=\{\theta(z|\e^a),\ z\ne0,\ 
2ilb\theta'(0|\e^a),\ z=0\}$, and only nonsingular terms contribute.
These terms are obtained when taking
the limit $b\to0$ in both 
arguments of the function $\theta(z|q)$:
\bea
\Bigl.\theta_{11}(z-2ilb|\e^{a+ib})\Bigr|_{b\to0}&=&
\sum_{n\in \bf Z}^{}\e^{-\frac{1}{2}(a+ib)(n+1/2)^2+(z-2ilb+i\pi)(n+1/2)}=
\nonumber\\
&=&\left[1-2ilb\frac{\d}{\d z}-i\frac{b}{2}\frac{\d^2}{\d z^2}
+O(b^2)\right]\theta_{11}(z|\e^a).
\label{d8}
\eea
This formula is valid everywhere except the zero point, at which we obtain
\bea
\Bigl.\theta_{11}(-2ilb|\e^{a+ib})\Bigr|_{b\to0}&=&
-2ilb\theta'_{11}(0|\e^a)-lb^2\theta'''_{11}(0|\e^a)+O(b^3),
\label{d9.1}
\\
\Bigl.\theta'_{11}(0|\e^{a+ib})\Bigr|_{b\to0}&=&
\theta'_{11}(0|\e^a)-\frac{ib}{2}\theta'''_{11}(0|\e^a)+O(b^2).
\label{d9.2}
\eea
Note that $\theta^{(2n)}(0|q)\equiv0$. 
Therefore, we obtain
\bea
&{}&\hbox{Re\,}\sum_{l=1}^{\infty}\left[\hbox{[local]\,}
-4a\e^{2il\vph}{
\bigl[1-2ilb\d-i\frac{b}{2}\d^2\bigr]\theta(\tau+i\vph|\e^a)
\bigl[\d-i\frac{b}{2}\d^3\bigr]\theta(0|\e^a)\over
\bigl[1-i\frac{b}{2}\d^2\bigr]\theta(\tau+i\vph|\e^a)
\bigl[-2ilb\d -lb^2\d^3\bigr]\theta(0|\e^a)}
+4a\{l\to -l\}\right]=\nonumber\\
&{}&\qquad=\bigl[\hbox{keeping\ only\ the\ part\ independent\ on\ $b$}\bigr]
\nonumber\\
&{}&\qquad=\hbox{Re\,}\sum_{l=1}^{\infty}\left[\hbox{[local]\,}
-4a\e^{2il\vph}\frac{\d}{\d z}\log\theta(\tau+i\vph|\e^a)+4a\{l\to-l\}\right].
\label{d10}
\eea
The remaining sum in $l$ can be easily done:
\be
\sum_{l=1}^{\infty}(\e^{2il\vph}-\e^{-2il\vph})=i\cot\vph,
\label{d11}
\ee
which yields the nonlocal contribution to the Green's function 
$G(\tau,\vph)$,
\be
G(\tau,\vph)=\frac{1}{4}\d_z\d_{\ov z}(-4a)\hbox{Re\,}
\left[i\cot\vph\frac{\d}{\d z}\log\theta(\tau+i\vph|\e^a)\right].
\label{d12}
\ee

Exploiting the standard product formula for the function 
$\theta_{11}(z|q)$,
$$
\theta_{11}(z|q)=ic(q)[\e^{z/2}-\e^{-z/2}]\prod_{m>0}^{}
(1-q^m\e^z)(1-q^m\e^{-z}),
$$
we find
\be
G(\tau,\vph)=\frac{1}{4}\d_z\d_{\ov z}(4a)\cot\vph
\hbox{Im\,}\left\{\frac{1}{\e^z-1}
+\sum_{m=1}^{\infty}\left[\frac{1}{q^m\e^z-1}-\frac{1}{q^m\e^{-z}-1}\right]
\right\}.
\label{d13}
\ee
Since 
\bea
\frac{\cos\vph}{\sin\vph}\hbox{Im\,}\frac{1}{q^m\e^z-1}&=&
\frac{-\e^{\tau+am}\cos\vph}{\e^{2(\tau+am)}-2\e^{\tau+am}\cos\vph+1}
\nonumber\\
&=&\hbox{Re\,}\frac{-\e^{z+am}}{(\e^{z+am}-1)(\e^{{\ov z}+am}-1)},
\label{d14}
\eea
disregarding local contributions, we obtain
\bea
G(\tau,\vph)&=&a\d_z\d_{\ov z}
\hbox{Re\,}\left\{-\frac{1}{(\e^z-1)(\e^{\ov z}-1)}\right.-\nonumber\\
&{}&\qquad-\left.\sum_{m=1}^{\infty}\left[
\frac{\e^{z+am}}{(\e^{z+am}-1)(\e^{{\ov z}+am}-1)}
+\frac{\e^{am-z}}{(\e^{am-z}-1)(\e^{am-{\ov z}}-1)}\right]\right\}.
\label{d15}
\eea
The differentiation w.r.t.\ $z$ and~$\ov z$ eliminates
pure holomorphic and antiholomorphic parts, 
and the remaining parts constitute the single expression
\be
G(z,\ov z)=-\frac{a}{4}\sum_{m=-\infty}^{\infty}
\frac{1}{\sinh^2\frac{z+am}{2}\sinh^2\frac{\ov z+am}{2}}.
\label{Green}
\ee
This is the Green's function of two Yang--Mills tensor
field insertions on torus. At sigularity points, it has the proper
scaling behavior $G(r)\sim1/r^4$.

\newsection{Discussion}

For the case of rectangular torus, we have obtained the proper
Green's function expression (\ref{Green}) for the two-point correlator
of the Yang--Mills tensor insertions. This demonstrates again that
the AdS/CFT correspondence holds in our case where no singularity at
the AdS time infinity is assumed. A more interesting (but far more
involved technically) is the problem of verifying this correspondence
in actual gravitational calculations of a four-point correlation
functions. 
But even concerning the free-field theory, there remain
questions on the mass spectrum, on the generalization to nonrectangular
tori, to higher dimensions, etc. Of special interest is the question
whether it is possible to consider filled Riemann surfaces of higher genera.
The construction works well in this case, but the 
$\eps$-regularized surface, or the boundary of the integation domain, 
cannot be described in the invariant
distance terms (the structure of the set
of closed geodesics becomes very involved already starting from genus two);
however, we hope that one can obtain a proper answer using 
an approximation technique. In the present calculations, we 
disregarded all local contributions. However, these contributions 
can be important when considering additional boundary terms in the 
initial action. It would be interesting to check whether the Hamiltonian
prescription of~\cite{ArFr} holds in this case.
These questions deserve further investigation.

\newsection{Acknowledgements}
The author thanks M.~A.~Olshanetsky for the valuable remark. The work
was supported by the Russian Foundation for Basic Research (Grant 
No.\,98--01--00327).

\end{document}